\documentstyle[PASJadd]{PASJ95}
\newcommand{\vol}{52}
\newcommand{\no}{4}
\newcommand{\lett}{}
\newcommand{\spage}{1}
\newcommand{\rdate}{2000 January 27}
\newcommand{\adate}{2000 March 16}

\begin{document}
\setcounter{page}{1}

\title{ISO Observations of the Planetary Nebula Lindsay 305 in the Small 
Magellanic Cloud\thanks{Based on observations with ISO, an ESA project with 
instruments funded by ESA Member States (especially the PI countries: 
France, Germany, the Netherlands, and the United Kingdom) and with the 
participation of ISAS and NASA.}}

\author{ Ar\=unas {\sc Ku\v{c}inskas}\thanks{Research Fellow of the Japan 
Society for the Promotion of Science.},$^1$ $^2$ $^3$  
Vladas {\sc Vansevi\v{c}ius},$^4$  
Marc {\sc Sauvage},$^5$ and 
Toshihiko {\sc Tanab\'{e}}$^6$
\\[12pt]
$^1${\it National Astronomical Observatory, Tokyo 181-8588, Japan}
\\
$^2${\it Institute of Theoretical Physics and Astronomy, Vilnius 2600, 
Lithuania}
\\
$^3${\it Institute of Material Research and Applied Science, Vilnius 
University, 
Vilnius 2009, Lithuania}
\\
$^4${\it Institute of Physics, Vilnius 2600, Lithuania}
\\
$^5${\it CEA/DSM/DAPNIA/Service d'Astrophys. C. E. Saclay, F-91191 
Gif-sur-Yvette Cedex, France}
\\
$^6${\it Institute of Astronomy, School of Science, The University of 
Tokyo, Tokyo 181-0015, Japan}
\\
{\it  E-mail(AK): arunaskc@cc.nao.ac.jp}}

\abst
{We present ISO (Infrared Space Observatory) observations of the planetary 
nebula Lindsay 305 (L 305) in the Small Magellanic Cloud.  
L 305 is very prominent in the ISOCAM frames at 6.75 and 11.5 $\mu$m, 
although it is under the detection limit at 4.5 $\mu$m.  
The obtained spectral energy distribution shows a strong mid-IR excess, 
which, depending on the amount of energy radiated at wavelengths longer 
than 11.5 $\mu$m, may be as large as $\sim 1500 L_{\odot}$.  
However, since an accurate estimate of the total nebular luminosity is not 
available to date, the evolutionary status of L 305 can not yet be 
constrained precisely.}

\kword{space vehicles: ISO satellite --- infrared: stars --- Magellanic 
Clouds: SMC --- planetary nebulae: individual (L 305) --- stars: 
evolution --- stars: AGB and post-AGB}

\maketitle
\thispagestyle{headings}

\section
{Introduction}

Lindsay 305 (L 305, Lindsay 1961; SMP 21, Sanduleak et al. 1978) is a 
planetary nebula located in the vicinity of the young populous cluster 
NGC 330 in the Small Magellanic Cloud (SMC).  
The optical spectrum of L 305 displays abundant forbidden emission lines 
typical of planetary nebulae ([O\,{\sc III}], [Ne\,{\sc V}], 
[Ne\,{\sc III}], [N\,{\sc II}] etc. - see e.g. Monk et al. 1988).  
Monk et al. (1988) and Leisy and Dennefeld (1996) have classified L 305 
as a Type I planetary nebula, whereas the central object of the nebula 
has been suspected to be a binary system (Leisy, Dennefeld 1996).  
Recent Hubble Space Telescope (HST) Faint Object Camera (FOC) imaging 
has revealed a noticeable asymmetry of L 305 (Vassiliadis et al. 1998, 
hereafter V98; Stanghellini et al. 1999, hereafter S99).

The evolutionary status of L 305 is rather unclear.  
The dynamical age estimates obtained by V98 and S99 from HST imaging 
differ by more than a factor of 5, and hence do not allow one to set 
out a reliable age estimate of L 305 (see section 3).  
However, if the luminosity and effective temperature of the central star 
of the nebula is known, a comparison of these parameters with the 
theoretical models of post-AGB evolution may yield an independent 
estimate of the nebular age.

It is well known, however, that planetary nebulae are surrounded by dust 
envelopes, which produce strong excess at the mid-infrared (mid-IR) 
wavelengths (e.g., Pottasch 1997).  
Any reliable luminosity estimate should therefore account adequately for  
the amount of energy emitted in the infrared.  
Thus, the mid-IR observations would help to constrain the total amount of 
energy emitted by the nebula (and thus the luminosity of the central star) 
and would allow one to further clarify the evolutionary status of L 305.

In this work we present ISOCAM observations of L 305.  
Since no previous observations of L 305 at mid-IR wavelengths are available, 
our data offers the first possibility of probing the dusty envelope of this 
nebula.  
Combining our data with the observations available from the literature we 
briefly discuss the evolutionary status of L 305.

\section
{Observations and Results}

The planetary nebula L 305 was observed during the course of raster 
imaging observations of the populous cluster NGC 330 with the ISOCAM 
(Cesarsky et al. 1996) on board the ISO satellite (Kessler et al. 1996).  
Observations were made on 1997 May 22 using broad-band CAM filters (LW1, 
LW2, and LW10, corresponding to the effective wavelengths of 4.5, 6.75, 
and 11.5 $\mu$m, respectively) with a pixel field of view (PFOV) of 
3$^{\prime\prime}$.  
The raster mode was 5 $\times$ 5, with the raster step size equal to 8 
pixels (24$^{\prime\prime}$).  
The fundamental integration time was set to $t_{\rm int}$ = 2.1 s, with 
a total number of 15 exposures per single raster position.

ISOCAM data were reduced using CAM Interactive Analysis software, CIA 
version 3.0 (The ISOCAM data presented in this paper was analyzed using 
``CIA'', a joint development by the ESA Astrophysics Division and the 
ISOCAM Consortium led by the ISOCAM PI, C. Cesarsky, Direction des 
Sciences de la Mati\`{e}re, C. E. A., France).  
Photometry was performed with the IRAF APPHOT package.  
The obtained ISOCAM fluxes are equal to 1.2 mJy and 5.5 mJy at LW2 and 
LW10, respectively.  
Since L 305 was not detected at 4.5 $\mu$m, our data yields only an 
upper limit estimate of $<$0.5 mJy at this wavelength.  
The absolute photometric uncertainty of our measurements comes mainly 
from the ISOCAM calibration uncertainties and a correction for the memory 
effect (transient) of the ISOCAM, which is estimated to be less than 20\% 
(Biviano 1998). 

The optical counterpart of the infrared source was identified from the 
instrumental coordinates of L 305, which were derived with respect to the 
relative positions of 8 field stars on LW10 CAM frame (identification 
accuracy is $\sim$ 1$^{\prime\prime}$).  
An optical identification chart of L 305 is shown in figure 1.



\section
{Discussion}

Currently available observational and theoretical data provide quite a 
diverse view of the evolutionary status of L 305.  
Photoionization models (Dopita, Maetheringham 1991) give $T_{\rm eff} = 
95\,000$ K and $L_{\ast} = 1750 L_{\odot}$ for the central ionizing source.  
Together with the theoretical tracks of post-AGB evolution, these 
parameters yield a main-sequence mass of $1.0 M_{\odot}$ and $1.4 
M_{\odot}$ for the He and H burning central star, respectively 
(Vassiliadis, Wood 1994).  
A direct estimate of dynamical age of the envelope (based on the 
[O\,{\sc III}] $\lambda$ 500.7 nm narrow-band images taken with the FOC on 
board the HST, and expansion velocities derived from the spectral 
observations of the same line) gives $\tau_{\rm dyn}$ = 21\,400 yr (S99).  
These facts quite consistently indicate that L 305 should be an evolved 
planetary nebula. 

However, at least two arguments work against the evolved dynamical status 
of L 305.  
First, the dynamical age derived by V98 (obtained using the same HST FOC 
image and [O\,{\sc III}] spectral data as in S99) is much smaller, i.e., 
$\tau_{\rm dyn} \sim$ 2500 yr and $\tau_{\rm dyn} \sim$ 4000 yr for the 
He-burning and H-burning central star scenarios, respectively.  
Second, the high electron density ($4\times 10^{4}$ cm$^{-3}$, Monk et al. 
1988; $10^{4}$ cm$^{-3}$, Leisy, Dennefeld 1996) and small nebular size of 
L 305 also suggest a young dynamical age. 

Indeed, an estimate of the dynamical age is strongly dependent upon the 
nebular geometry, the method used to determine the nebular expansion 
velocity and so forth.  
The expansion velocity, as measured by V98, was defined at the 10\% level 
of the maximum line intensity, and therefore probed the fastest moving 
material of the nebula (yielding the expansion velocity of 
$v_{\rm exp}=$35.2  km s$^{-1}$).  
S99, however, has defined the expansion velocity at the full width at 
half-maximum (FWHM) of the line, obtaining $v_{\rm exp}=19.3$ km s$^{-1}$.  
Therefore, the dynamical age obtained by V98 should be systematically 
smaller than that obtained by S98 by a factor of $\sim 2$.  
This, however, still can not account for the surprisingly large difference 
between the dynamical ages derived by S99 and V98.  
One of the possibilities to discriminate between these age estimates is to 
make a comparison between the evolutionary parameters of L 305 derived 
from the dynamical age estimates mentioned above and those obtained by 
other methods.

If the effective temperature of the central object is known, the dynamical 
age combined with the theoretical models of stellar evolution can provide 
a luminosity estimate of the central star.  
Employing theoretical models of the post-AGB evolution of Vassiliadis and 
Wood (1994) and assuming that the effective temperature of the central 
object is 95\,000 K (Dopita, Maetheringham 1991) one obtains $\sim$ 8000 
$L_{\odot}$ and $\sim$ 3500 $L_{\odot}$ for the dynamical ages of V98 
and S99, respectively.  
Both of these estimates are considerably larger than 1750$ L_{\odot}$ 
predicted by Dopita and Maetheringham (1991) from the photoionization 
models.

The discrepancy between these luminosity estimates may be due to the fact 
that the luminosity of the central object derived by Dopita and 
Maetheringham (1991) was obtained assuming that the extinction towards 
L 305 was equal to zero. 
However, at least several independent studies show that the extinction 
towards L 305 is non-negligible.  
The reddening constant at H$\beta$, $c$(H$\beta$) (the definition of which 
comes from $I_{\rm cor}(\lambda)/I_{\rm cor}({\rm H}\beta) = I_{\rm obs}
(\lambda)/I_{\rm obs}({\rm H}\beta)\,10^{c({\rm H}\beta)\,f(\lambda)}$, 
where $I_{\rm obs}(\lambda)$ and $I_{\rm cor}(\lambda)$ are observed and 
corrected intensities at wavelength $\lambda$ and $f(\lambda)$ is the 
reddening curve, as parametrized by Miller and Mathews, 1972) was found 
to be 0.79 and 0.44 by Monk et al. (1988) and Leisy and Dennefeld (1996), 
respectively.  
If we crudely assume that the flux in H$\beta$ is directly proportional 
to stellar luminosity (e.g., Kaler 1976) and take the averaged reddening 
constant to be $\sim$ 0.6, we will find that the luminosity of the central 
object in L 305 may be underestimated up to by a factor of $\sim$ 4, if 
the reddening is neglected.  
The corrected luminosity ($\sim 7000 L_{\odot}$) would imply an 
evolutionary age of $\sim$ 4200 yr for the H-burning central star 
(Vassiliadis, Wood 1994), which would be in good agreement with the 
direct estimate obtained by V98 ($\sim$ 4000 yr).


The interstellar reddening towards NGC 330 is small and covers the range 
from $E(B-V)=0.03$ (Carney et al. 1985) to $E(B-V)=0.12$ (Bessell 1991). 
Therefore, the high reddening discussed by Monk et al. (1988) and Leisy 
and Dennefeld (1996) should be of circumstellar origin.  
However, if the reddening is caused by circumstellar extinction in the 
dust envelope surrounding the nebula, the absorbed radiation should be 
re-emitted at longer wavelengths.  
Indeed, the spectral energy distribution of L 305 (figure 2) shows a 
prominent excess in the mid-infrared ($\lambda \ge 7 \mu$m), which can 
be due to thermal emission by circumstellar dust.  
A simple estimate (obtained by integration over the blackbody fit of 
the observed mid-infrared fluxes) shows that the amount of energy 
emitted in the infrared is $L_{\rm IR} \sim 250 L_{\odot}$.  
This is obviously too low to account for the difference between 1750 
$L_{\odot}$ obtained by Dopita and Maetheringham (1991) and $\sim 7000 
L_{\odot}$ predicted from the reddening considerations above.

Indeed, at the dust temperatures typical for the planetary nebulae, 
$\sim$ 100--250 K (see e.g. Tajitsu, Tamura 1998), the largest fraction 
of energy will be re-emitted by dust at wavelengths longer than $\sim 
12 \mu$m. 
Using a sample of planetary nebulae studied by Zhang and Kwok (1991) 
as an example, we estimate that the total amount of energy emitted in 
the infrared is typically 4--5 times greater than the amount of energy 
estimated from the blackbody fit to the mid-IR data 
($\lambda =$ 7--12 $\mu$m).  
Thus, the total infrared luminosity of L 305 could be considerably larger, 
i.e., reaching $L_{\rm IR} \sim$ 1000--1500 $L_{\odot}$.  
Together with $\sim 1750 L_{\odot}$ obtained by Dopita and Maetheringham 
(1991), this would come close to $\sim 3500 L_{\odot}$ (implied from the 
dynamical age estimate of S99), giving support for a low-mass evolved 
nebula scenario.

It should be stressed, however, that due to the complicated geometries 
of the planetary nebulae, a considerable amount of stellar photons can 
escape without affecting the nebula, itself.  
In such a case, the total luminosity of L 305 would be considerably 
higher than the luminosity estimate obtained by Dopita and Maetheringham 
(1991) and corrected for the amount of energy re-radiated at the mid-IR 
wavelengths.  
However, the fraction of photons escaping the nebula is not known in the 
case of L 305.  
Therefore, in view of this argument (and other contradictory facts 
mentioned above), it still seems impossible to derive a precise 
luminosity estimate for L 305.  
Hence, even though our results point to a low-mass evolved nebula 
scenario, the evolutionary status of L 305 still needs to be clarified 
by future observations made in the UV, near-IR, and far-IR wavelengths 
backed up by a comprehensive self-consistent theoretical model of L 305.

\section
{Conclusions}

We present ISOCAM observations of the planetary nebula L 305 located in 
the vicinity of the young populous cluster NGC 330 in the Small Magellanic 
Cloud.  
Previous observations of this object have given quite a diverse view on 
the evolutionary status of L 305, indicating that it may be either a 
dynamically young planetary nebula with a high-mass central object, or 
a rather evolved nebula with a low-mass central star.  
Our observations have revealed that L 305 shows a strong excess in the 
mid-IR.  
We estimate that, depending on the amount of radiation emitted at 
wavelengths longer than 11.5 $\mu$m, the total infrared luminosity can 
be as high as $\sim 1500$ $L_{\odot}$.  
However, the presently available observations do not allow us to derive 
a precise estimate of the total luminosity of the nebula.  
Therefore, even though our results would support a low-mass evolved 
nebula scenario, the evolutionary status of L 305 still remains to be 
constrained by future studies.

\par
\vspace{1pc}\par
We thank an anonymous referee for valuable comments and suggestions, and 
Yollande McLean for a careful reading of the manuscript. This research
was supported in part by grant-in-aids for Scientific Research (C) and 
for International Scientific Research (Joint Research) from the Ministry 
of Education, Science, Sports and Culture in Japan. 

\section*{References}
\small

\re
Balona L.A. 1992, MNRAS 256, 425

\re
Bessell M.S. 1991, in IAU Symp. 148, The Magellanic Clouds, ed 
R. Haynes, D. Milne (Kluwer, Dordrecht) p 273

\re
Biviano A. 1998, in The ISOCAM Calibration Error Budget Report, version 3.1, 
p 16

\re
Carney B.W., Janes K.A., Flower P.J. 1985, AJ 90, 1196

\re
Cesarsky C.J., Abergel A., Agnese P., Altieri B., Augu\`{e}res J.L., 
Aussel H., Biviano A., Blommaert J. et al. 1996, A\&A 315, L32

\re
Dopita M.A., Maetheringham S.J. 1991, ApJ 367, 115

\re
Kaler J.B. 1976, ApJ 210, 843

\re
Keller S.C., Wood P.R., Bessell M.S. 1999, A\&AS 134, 489

\re
Kessler M.F., Steinz J.A., Anderegg M.E., Clavel J., Drechsel G., Estaria P., 
Faelker J., Riedinger J.R. et al. 1996, A\&A 315, L27

\re
Leisy P., Dennefeld M. 1996, A\&AS 116, 95

\re
Lindsay E.M. 1961, AJ 66, 169

\re
Miller J.S., Mathews W.G. 1972, ApJ 172, 593

\re
Monk D.J., Barlow M.J., Clegg R.E.S. 1988, MNRAS 234, 583

\re
Pottasch S.R. 1997, in Planetary Nebulae, ed. H.J. Habing, H.J.G.L.M. Lamers 
(Kluwer, Dordrecht) p 483

\re
Sanduleak N., MacConnell D.J., Philip A.G.D. 1978, PASP 90, 621

\re
Stanghellini L., Blades J.C., Osmer S.J., Barlow M.J., Liu X.-W. 
1999, ApJ 510, 687 (S99)

\re 
Tajitsu A., Tamura S. 1998, AJ 115, 1989

\re 
Vallenari A., Ortolani S., Chiosi C. 1994, A\&AS 108, 571

\re 
Vassiliadis E., Wood P.R. 1994, ApJS 92, 125

\re 
Vassiliadis E., Dopita M.A., Meatheringham S.J., Bohlin R.C., Ford H.C., 
Harrington J.P., Wood P.R., Stecher T.P., Maran S.P. 1998, ApJ 503, 253 (V98)

\re 
Zhang C.Y., Kwok S. 1991, A\&A 250, 179

\label{last}

\newpage

\begin{table}
\small
\begin{center}
Table~1.\hspace{4pt}Optical photometry of L 305.\\
\end{center}
\vspace{6pt}
\begin{tabular*}{\columnwidth}{@{\hspace{\tabcolsep}
\extracolsep{\fill}}cccc} 
\hline\hline\\[-6pt]
 $B$    & $V$    & $I$    & Source \\         
[4pt]\hline\\[-6pt]
         & 17.44  & 18.25 & Keller et al. 1999    \\    
  17.61  & 17.33  &       & Vallenari et al. 1994 \\    
  17.83  & 17.28  &       & Balona 1992           \\    
\hline
\end{tabular*}
\end{table}


\vspace*{20mm}

\section*{Figure Captions}

\begin{fv}{1}{18pc}%
{Identification chart of L 305 in $I$-band (north is up and east is left).  
Part of the SMC populous cluster NGC 330 is seen at the lower right.  
The insert shows a HST FOC [O\,{\sc III}] $\lambda$ 500.7 nm image of L 305 
taken from V98 (scale is $1''$ on each side of the insert box).}
\end{fv}

\begin{fv}{2}{18pc}%
{Spectral energy distribution of L 305, constructed from optical photometry 
(table 1) and ISO data.  
The error bars of the mid-IR data represent formal IRAF/APPHOT errors.  
The solid line shows a blackbody fit to the ISOCAM data ($T_{\rm BB} = 
280$ K).} 
\end{fv}

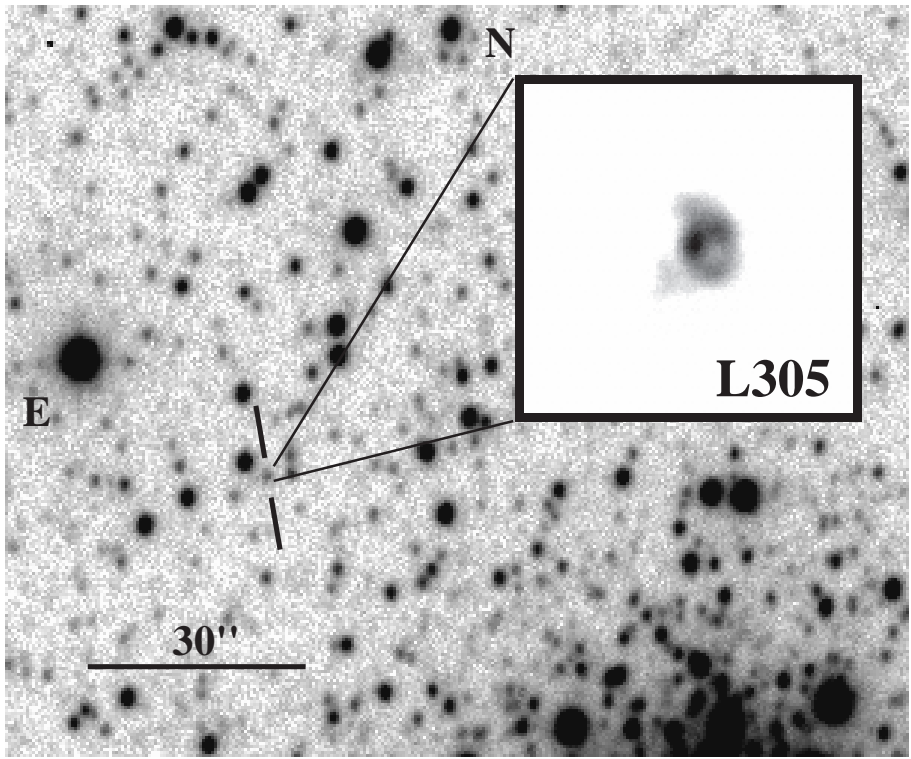
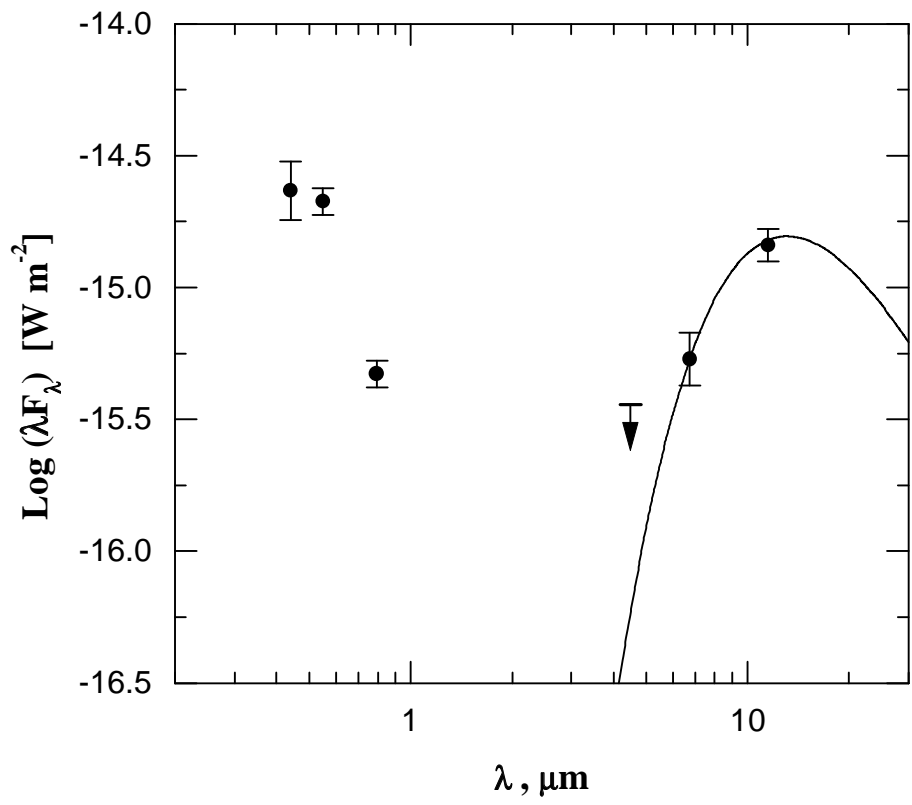
\end{document}